\documentclass[preprint,12pt]{aastex}

\def\etal{{et al.~}}
\shortauthors{Lee et al.}

\begin{document}

\title{Velocity distribution of collapsing starless cores,
         L694-2 and L1197}

\author{SEOK HO LEE\altaffilmark{1}, YONG-SUN PARK\altaffilmark{1}, JUNGJOO SOHN\altaffilmark{1}, CHANG WON LEE\altaffilmark{2}, AND HYUNG MOK LEE\altaffilmark{1}}

\altaffiltext{1}{Department of Physics and Astronomy, Seoul
National University, Shillim-dong Kwanak-gu, Seoul 151-742, Korea
} \altaffiltext{2}{Korea Astronomy and Space Science Institute,
61-1 Hwaam-dong, Yusung-gu, Taejon 305-348, Korea}
\begin{abstract}
  In an attempt to understand the dynamics of collapsing starless cores,
  we have conducted a detailed investigation of the velocity fields of two
  collapsing cores, L694-2 and L1197, with high spatial resolution HCN J~=~1-0
maps and Monte Carlo radiative transfer calculation.
  It is found that infall motion is most active in the middle and outer
layers outside the central density-flat region, while both the
central and outermost parts of the cores are static or exhibit
slower motion.
  Their peak velocities are 0.28~km~s$^{-1}$ for L694-2 and
0.20~km~s$^{-1}$ for L1197, which could not be found in simple
models.
  These velocity fields are roughly consistent with the gravitational
collapse models of the isothermal core; however, the velocity
gradients inside the peak velocity position are steeper than those
of the models.
  Our results also show that the density distributions are $\sim r^{-2.5}$ and $\sim  r^{-1.5}$
  in the outer part for L694-2 and L1197, respectively.  HCN abundance relative to H$_2$ is spatially
almost constant in L694-2 with a value of $7 \times 10^{-9}$,
while for L1197, it shows a slight inward increase from $1.7
\times 10^{-9}$ to $3.5 \times 10^{-9}$.
\end{abstract}
\keywords{ISM: individual (L694-2, L1197) --- ISM: kinematics and
dynamics ---star: formation}
\section{Introduction }
Starless cores that are on the verge of or undergoing
gravitational collapse  provide important information on the
initial stage of star formation. In order to understand the
dynamics of such starless cores, it is necessary to obtain
information on at least density, temperature, and velocity
distributions. The density and temperature have been determined by
submillimeter/NIR and NH$_3$ observation with a certain degree of
accuracy \citep{evans, kandori, tafalla2004}.

However, the velocity field has not yet been elucidated. The
observations of starless cores in the molecular lines of CS,
$\rm{DCO^{+}}$, and $\rm{N_{2}H^{+}}$
\citep{surveya,surveyb,surveyc} show that the inward motion is
extended to $0.2 - 0.3$ pc with a speed of
$0.05-0.09$~km~s$^{-1}$, which is derived  using a two-layer model
\citep{myersetal}.
  The analysis of the CS J~=~3-2 and J~=~2-1 lines with the same model for several
cores such as L183, L1521F, L1689B, L1544, and L694-2, suggests
that the infall speed appears to increase toward the center
\citep{surveyc}. Since the two layer model is rather simple, the
derived velocity may be an average value along the line of sight
and its variation with depth should be considered with caution.
Moreover, it should be noted that in general, CS is significantly
depleted in the central regions of starless cores
\citep{tafalla2002, tafalla2004}. On the other hands, by applying
an improved version of the two layer model \citep{devries} to the
interferometric observations in $\rm{N_{2}H^{+}}$,
\citet{williams} found that the infall velocity increases toward
the emission peak for L694-2 and L1544 .
    This transition might trace the deep central
region of the core because the emission of $\rm{N_{2}H^{+}}$
primarily comes from the central part due to its low optical depth
and centrally condensed matter distribution \citep{tafalla2006}.
Therefore, it is highly likely that these independent observations
do not trace the entire range of the velocity fields in the
collapsing cores. Similar shortcomings are evident in
\citet{tafalla2002}, \citet{keto}, and \citet{swift}, although
they have assumed that the infall speed varies monotonically with
the radial distance and have performed more sophisticated
radiative transfer calculations.

  In order to derive a reliable velocity field, by overcoming these problems,
we selected HCN J~=~1-0 hyperfine lines as a probe and observed
several candidates of collapsing cores.
  Then, we performed radiative transfer calculations by considering different types of  velocity fields.
  An advantage of using the HCN J~=~1-0 lines is that three hyperfine
lines are well separated and they cover a wide range (a factor of
5) of the optical depth.  In addition, a direct comparison between
the integrated intensity map of HCN and that of the
$\rm{N_{2}H^{+}}$
  reveals that HCN is not depleted significantly \citep{sohn}.
  For radiative transfer analyses, we used the one-dimensional Monte Carlo
code by considering the line overlap effect due to the hyperfine
splitting of HCN energy levels \citep{bernes, gonzalez, park}.
  Therefore, the capability of this code is almost the same as that of code used by
\citet{tafalla2002}, \citet{keto}, and \citet{swift}.
  The only difference is that we explored more complicated
velocity fields, by relaxing the condition of the monotonic
increase or decrease of velocity field, which is an implicit
assumption of the previous studies.

\section{Data}
  We analyzed two round-shaped collapsing starless cores, L694-2 and L1197, located at
$\alpha_{2000}=19^h41^m04.^s5$,
$\delta_{2000}=10^{\circ}57'02.''0$ and $\alpha_{2000}=
22^h37^m02.^s3$, $\delta_{2000}=58^{\circ}57'20.''6$,
respectively. The observations were performed in September 2003 by
using the IRAM 30 m at a spacing of 11$''$ \citep{sohn}.
  The beam size of the telescope was $\rm{FWHM}=28''$ for the HCN J~=~1-0 and the
velocity resolution of the backend was 0.033~km~s$^{-1}$. Data
were converted to the T$_{mb}$ scale using a main beam efficiency
of 0.82.  We set the frequency of the three HCN J~=~1-0  hyperfine
lines according to the Cologne Database for Molecular Spectroscopy
\citep{muller} as follows: 88.6304157~GHz for F~=~0-1,
88.6318473~GHz for F~=~2-1, and 88.6339360~GHz for F~=~1-1.

 In order to perform a comparison
with the one-dimensional radiative transfer calculations, we
assumed that two cores are spherically symmetric. It is justified
by the spherically symmetric density distribution from the
observation of dust extinction and emission  for L694-2
\citep{harveya, harveyb}. The HCN J~=~1-0 line profiles are
distributed concentrically and they all exhibit blue asymmetry for
L694-2. This is the case for L1197 except for red asymmetry in
south-west region occupying only $\le$ 10\% of the observed area,
which has a negligible contribution to the analysis.
  The observations will be described in detail in a separate paper
(Sohn \etal, in preparation). The observed spectra were
azimuthally averaged at concentric radial annuli with a step of
10$''$.  We set the centers of the cores to the peak intensity
positions of $\rm{N_{2}H^{+}}$, which are (0$''$, 0$''$) for
L694-2 and (0$''$, 11$''$) for L1197 relative to the map centers
mentioned above.
  Since the spectra were at rectangular grid points, they were weighted by
a factor of $exp(-(d/14'')^{2})$, where $d$ is the distance in
arcseconds between the radial distance of the grid point and the
specified radius and 14$''$ is one half width of telescope beam.
  The resulting spectra are shown in Fig.~\ref{fig1} and will be
used as templates for the comparison with synthesized line
profiles.

\section{Model of core}
  We synthesized line profiles by using the Monte Carlo radiative
transfer model and compared them with the observed ones in order
to derive the model parameters that best explain the observations.
  We adopted distances of 250pc and 400 pc to L694-2 and L1197, respectively \citep{surveyb}.
  Velocities relative to the local standard of rest were selected as
9.61~km~s$^{-1}$ and $-3.16$~km~s$^{-1}$ for L694-2 and L1197,
respectively.
  The model cores with a radius of 0.15 pc for both L694-2 and L1197 were composed of 30
concentric shells at regular intervals.
  Additionally, these cores were surrounded by an envelope with an adjustable
radius (see below). We used $10^6$ model photons for the Monte
Carlo code.

 The kinetic temperature was assumed to be spatially
constant at 10K, on the basis of the values derived for L1498 and
L1517B \citep{tafalla2004}.
  For the density distribution, we used the following form adopted by
\citet{tafalla2004},
\begin{eqnarray} \label{model} n(r)=
{n_{c}\over 1+(r/r_{o})^{\alpha}} \ \ ,
\end{eqnarray}
where $n_{c}$ is the central density, $r_{o}$ is the radius of the
inner flat region, and $\alpha$ is the asymptotic power index.
  In dynamic models, $r_{o}$ decreases with an increase in the
  central density \citep{foster, ciolek}.
  To consider this effect, we defined $r_o$ such that it depended on $n_c$,
\begin{eqnarray}
\label{model1} r_{o}\equiv2.25R_{o} \ \ ,\\
 R_{o}=a/\sqrt{4\pi G \mu m_{H} n_{c}} \ \ ,
\end{eqnarray}
where $R_{o}$ is the scale radius of the Bonnor-Ebert sphere, $a$
is the sound speed, $\mu$ (= 2.33) is the mean molecular weight,
and $m_{H}$ is the hydrogen mass.
  If $\alpha$ is 2.5 in equation (\ref{model}), it represents approximately
the density distribution of the Bonnor-Ebert sphere
\citep{tafalla2004}. In our model, we tested two cases of sound
speed, $a$ = 0.2 and 0.3~km~s$^{-1}$, and finally adopted the
latter.

  Since the integrated intensity map of the thinnest HCN J~=~1-0, F~=~0-1 line
is similar to that of the N$_2$H$^+$~J~=~1-0 line, it appears that
HCN is not significantly depleted for these two cores at least
\citep{sohn}.
  In order to take into account any minor depletion or enhancement of HCN, we assumed
the HCN abundance variation of $X(r)$~$=$~$X_{o}[n(r)/n_o]^\beta$,
where $\beta$ is the power index and $n_o$ and $X_o$ are the
density and HCN abundance relative to H$_2$ near the boundary,
respectively. $n_o$ were fixed lastly as $5 \times 10^3$ cm$^{-3}$
for L694-2 and $1 \times 10^4$ cm$^{-3}$ for L1197, respectively.

The line profiles are most sensitive to the infall velocity field.
We characterized the velocity field as a $\Lambda$-shaped field.
(The reason why we adopted this shape will be discussed in next
section.) The innermost region within $R_i$ is static, and the
infall velocity increases with a linear function of the radial
distance toward the peak velocity position ($R_{m}$) with a
maximum magnitude of $V_{m}$, and then linearly decreases to the
outermost layer. We fixed the size of the outermost layer and its
velocity to $0.13 {\rm pc} \leq r \leq 0.15 {\rm pc}$ and a
constant, respectively. In fact, the size of the outermost layer
is determined on the basis of the first several trials. The
velocity of the outermost layer can be directly derived from the
line profile. The fitting procedure will be described in detail in
next section.

In summary, we have seven free parameters in total : $n_c$ and
$\alpha$ for density, $X_o$ and $\beta$ for abundance, and $R_i$,
$R_m$, and $V_m$ for velocity. We made reference to the values of
previous studies for density and abundance \citep{harveya,
harveyb, gonzalez}.

Besides these, a few auxiliary parameters were necessary to fit
the line profiles. The e-folding widths of the absorption
coefficient profile in the inner static region were specified as
0.15~km~s$^{-1}$ and 0.13~km~s$^{-1}$ for L694-2 and L1197,
respectively; they are larger than the pure thermal width by a
factor of $\sim 2$. In the other layers, these widths were
automatically adjusted such that the line widths of the model
spectra closely resembled the observed ones.
  To explain the hyperfine anomalies of HCN, an envelope with larger
microturbulence width and lower density must be introduced \citep
{gonzalez}.
  The diffuse and turbulent molecular clouds surrounding the cores traced by CO
may be considered as the envelope \citep{tafalla2006}.
  We selected the microturbulence width of 0.5 km~s$^{-1}$ and the abundance of
$5\times10^{-9}$, which were the same for both cores.
  The density was set to $10^3$~cm$^{-3}$ for L694-2 and $3 \times10^3$~cm$^{-3}$
for L1197.
  Then, we varied the size of the envelope to adjust the relative intensity ratio
among the hyperfine lines.  It is not necessary for the envelope
parameters to be unique since various combinations of the density,
abundance, and size are acceptable if the optical depth is almost
the same and the excitation temperature is sufficiently low and
close to 2.7~K. It should be noted that the envelope is introduced
to account for only the intensity ratio among the hyperfine lines.

The line profiles produced by this method were convolved with the
telescope beam for a direct comparison with the observations. The
difference between the model line profile and the observed one was
evaluated by $\chi^{2}$ \citep {zhou}. We considered all channels
of three hyperfine transitions on 5 positions, in the velocity
range from $-0.7$~km~s$^{-1}$ to $0.7$~km~s$^{-1}$.

\section{Results}

In the first phase, we adjusted the density distribution and
constant HCN abundance without infall motion so that the
intensities of F~=~0-1 at all locations are similar to those
observed. Further, we attempted to fit the line profiles with the
infall velocity fields of the constant or monotonically
increasing/decreasing function of the radius and found that they
cannot reproduce the observed spectra. The infall velocities at
every grid points were then varied by trial and error. It was
immediately evident that the infall motion is dominant in the
middle and outer layers and other parts of the cores are rather
static or exhibit a small amount of motion. Therefore, we
approximated the velocity field with the $\Lambda$-shaped one.

\begin{figure}
\epsscale{1.1} \plottwo{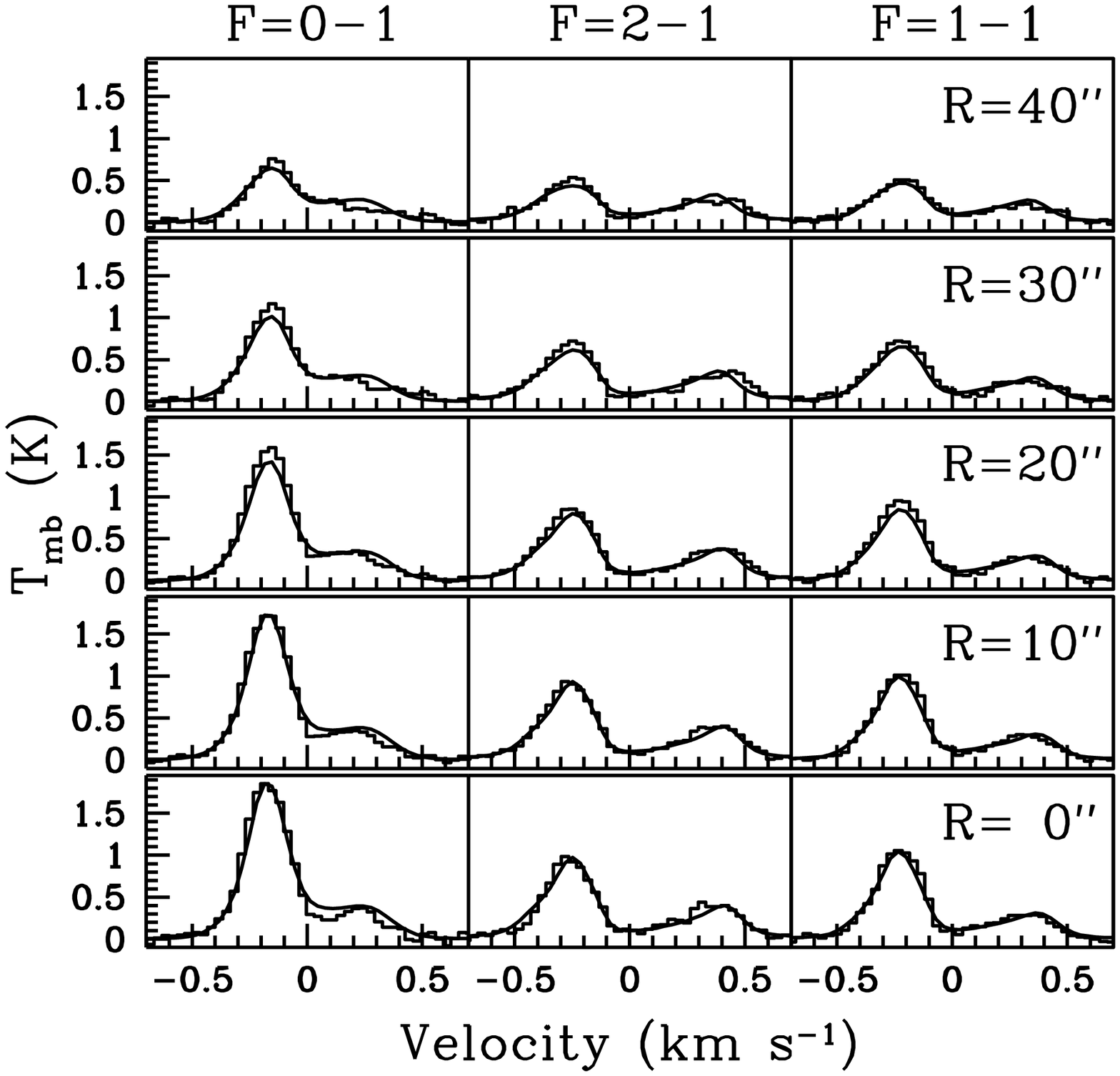}{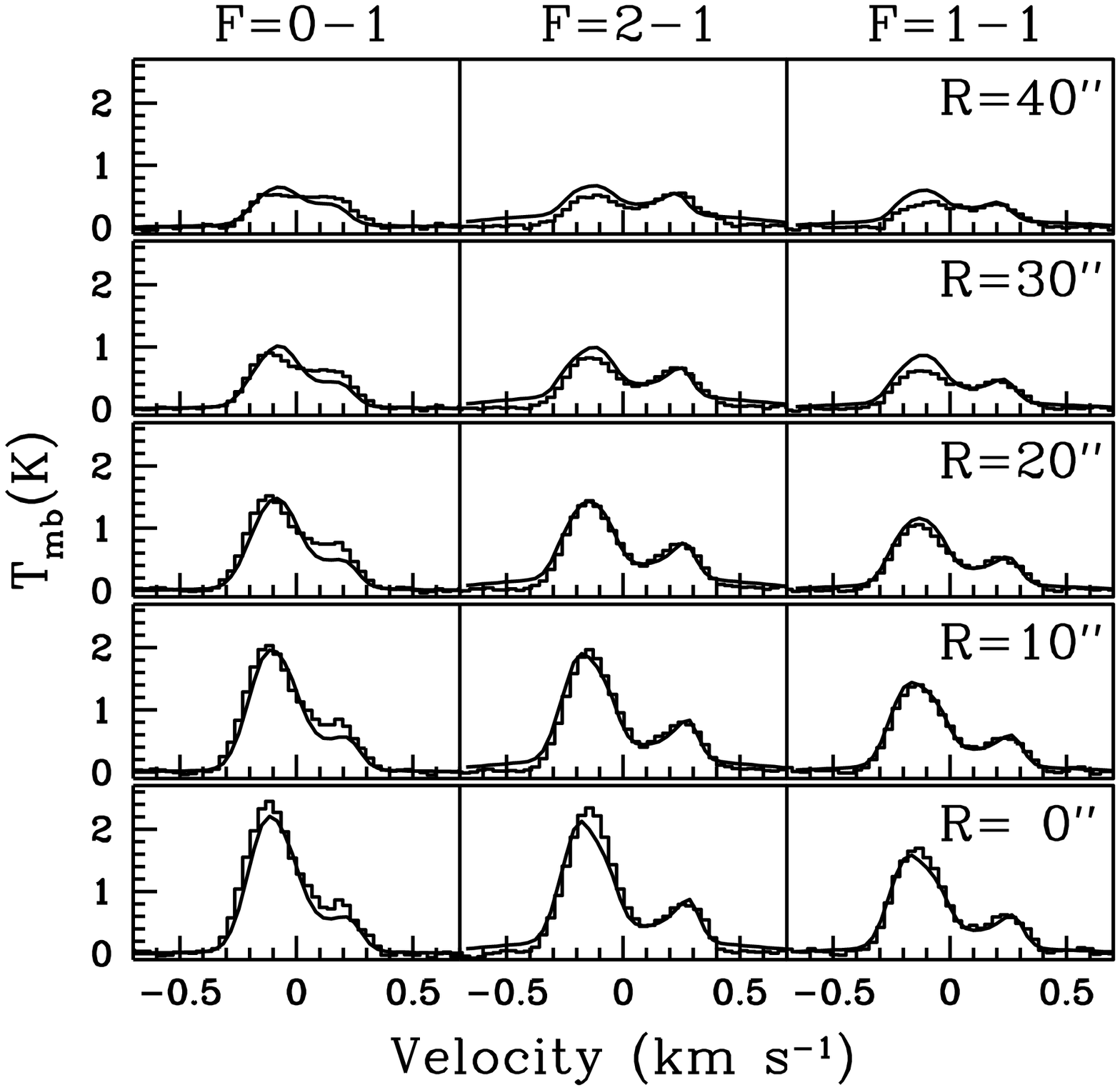}
\begin{center}
 \caption{The comparison of spectra from the best fit model (line)
and those from the observation (histogram) for L694-2 (left) and
L1197 (right) as a function of projected radial distance, which is
noted in the rightmost panels.
  Observed spectra are shifted so that their LSR velocity is
  zero. \label{fig1}}
\end{center}
\end{figure}

The $\Lambda$-shaped velocity field could be qualitatively
understood by inspecting the characteristic features of the
observed line profiles shown in Fig.~\ref{fig1} as follows. On the
basis of the fact that the position of the self-absorption dip in
the thickest F~=~2-1 line indicates the velocity at the outermost
part, we can estimate the infall velocity in that region. These
infall velocities for L694-2 and L1197 are $\approx 0.0$ and
$\approx 0.1$~km~s$^{-1}$, respectively. On the other hand, the
velocity in a region close to the central part can be derived from
the line width of an optically thin N$_2$H$^+$ line, since it is
not depleted significantly and, therefore, its emission mainly
originates from the central part in which the density is the
highest. From \citet{surveya}, the FWHM of L694-2 and L1197 are
found to be 0.27~km~s$^{-1}$ and 0.28~km~s$^{-1}$, respectively.
These values can be used to derive the contribution of non-thermal
motion after removing the thermal width.
  If this motion is attributed to the systematic inward motion, the maximum
infall velocity will be 0.12~km~s$^{-1}$ for both L694-2 and
L1197. Since cores with a pure thermal line width are extremely
rare, the systematic motion will be even smaller.
  The line shape also provides information on the velocity field in the middle layer.
  The increasing rate of the brightness from the absorption dip to red part is considerably slower
than to blue part, particularly for L694-2, thereby resulting in a
weaker red peak.
  This suggests that the region of $\tau_{\nu}=1$ remains close to the
surface layer in the red part, implying that the infall velocity
increases steeply toward the deeper layer.
  Then, it must decrease again toward the center to satisfy the constraints
of the N$_2$H$^+$ lines.

  In the second phase, in order to derive the velocity field in a more
systematic manner, we reduced the number of parameters describing
the velocity field to three, $R_i$, $R_m$, and $V_{m}$ as
mentioned in the previous section.
  First, we adjusted the parameters of the velocity field, while fixing the other density and
abundance parameters. Thereafter, the density parameters were
modified followed by the abundance parameters, while the other
parameters were unchanged. We repeated this process a few times.

\begin{figure}
\epsscale{1.1} \plottwo{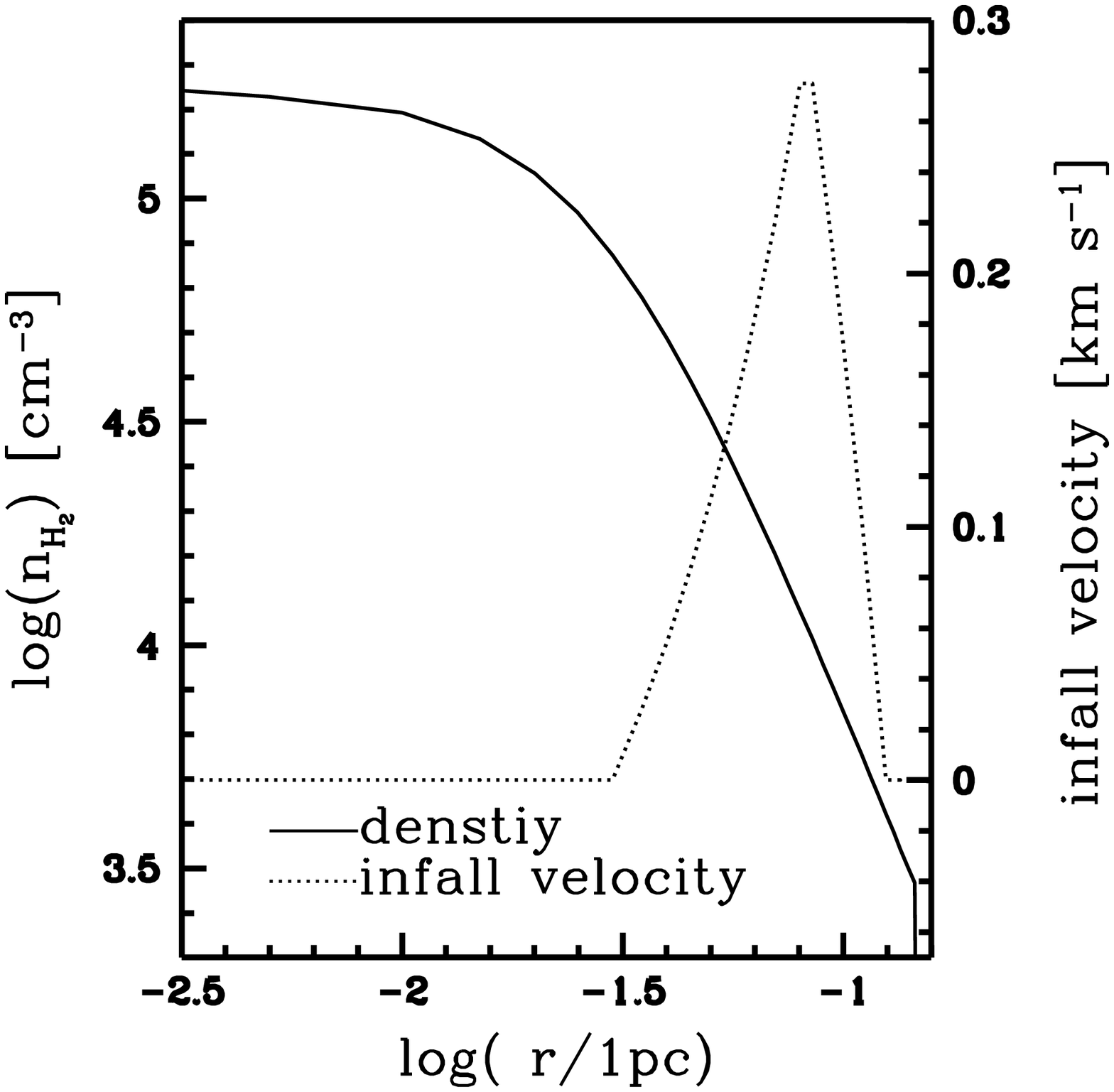}{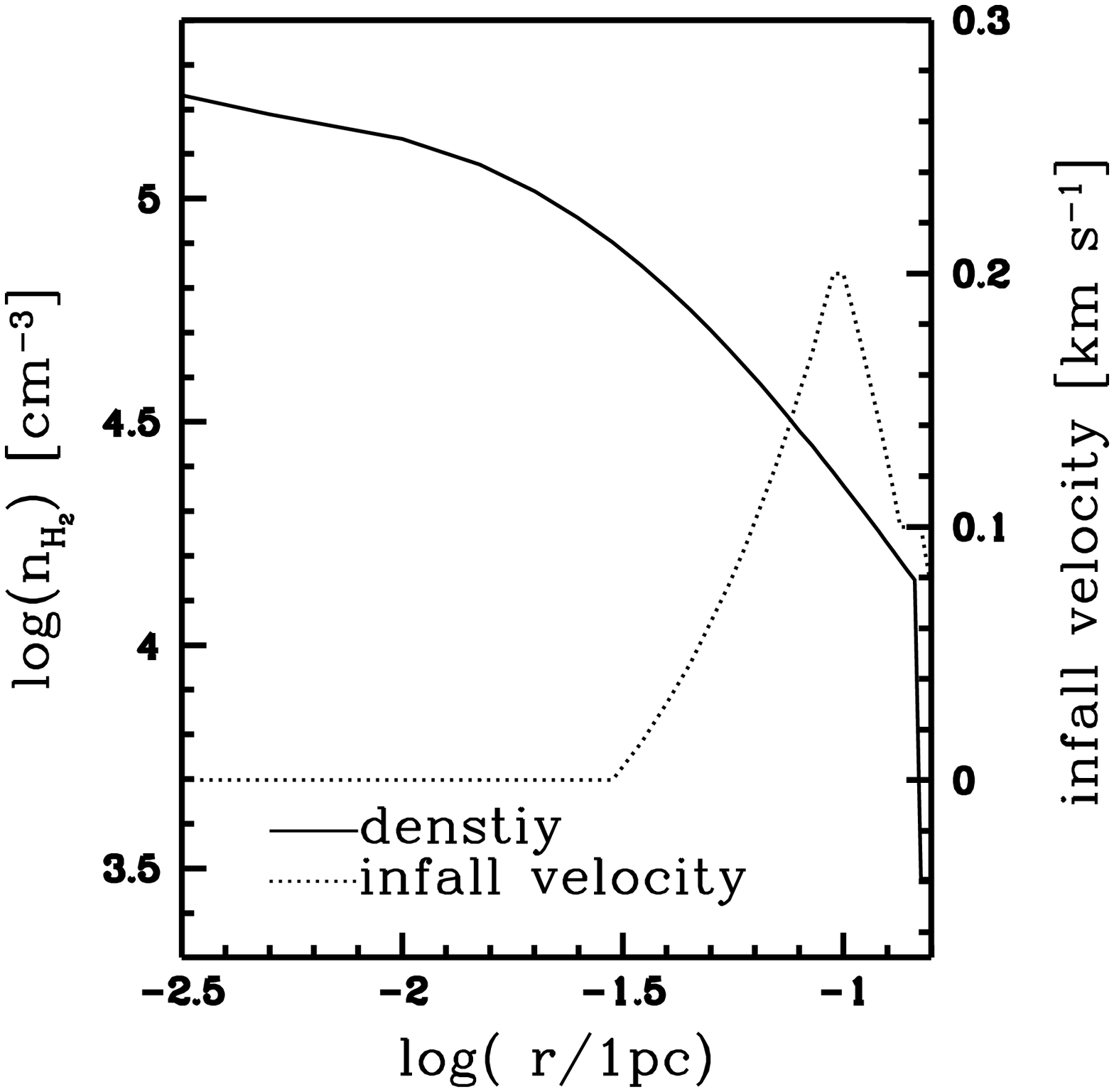}
\begin{center}
\caption{The density and infall velocity distributions of the best
fit model for L694-2 (left) and L1197 (right). \label{fig2}}
\end{center}
\end{figure}

\begin{figure}
\epsscale{1.0} \plotone{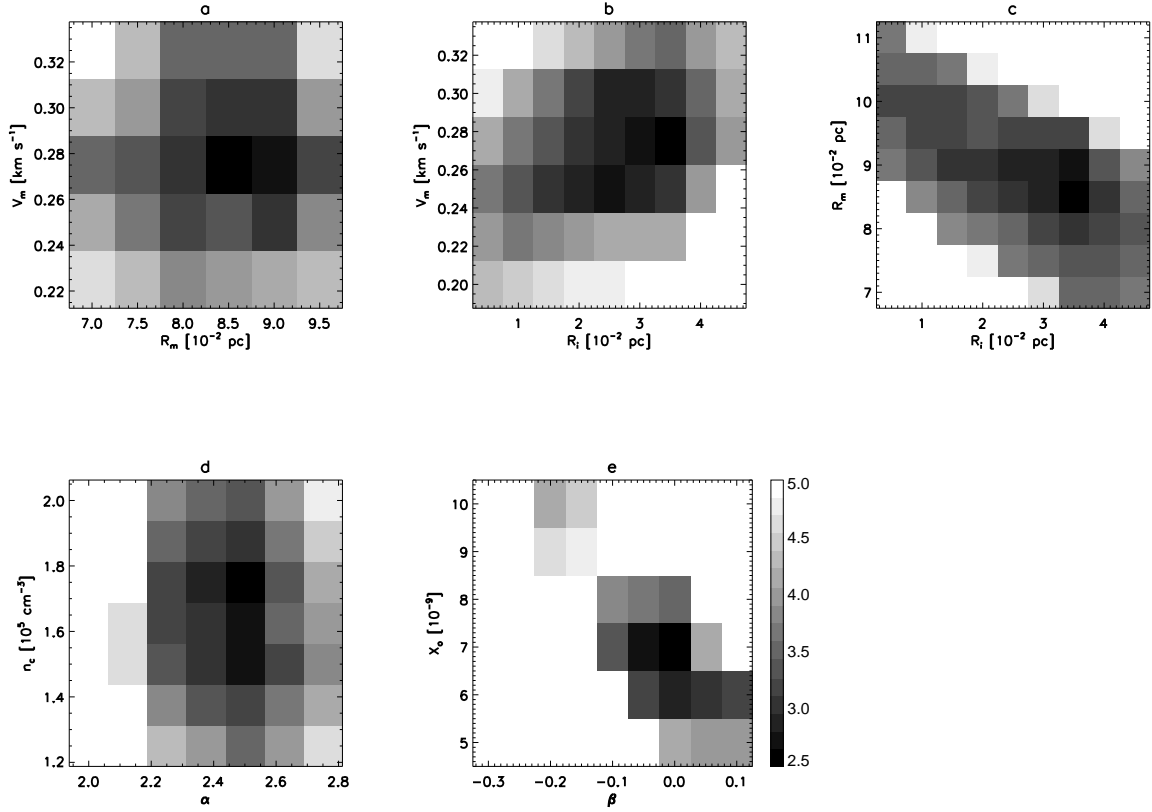}
\begin{center}
\caption{Distribution of $\chi^2$ around the best fit solution
with $\chi^2$ of 2.7 as a function of model parameters for L694-2
(see text for the best fit model parameters). Panels a, b, and c
are about velocity field, and d and e are about density and
abundance distribution, respectively.
  \label{fig3}}
\end{center}
\end{figure}

\begin{figure}
\epsscale{1.0}
 \plotone{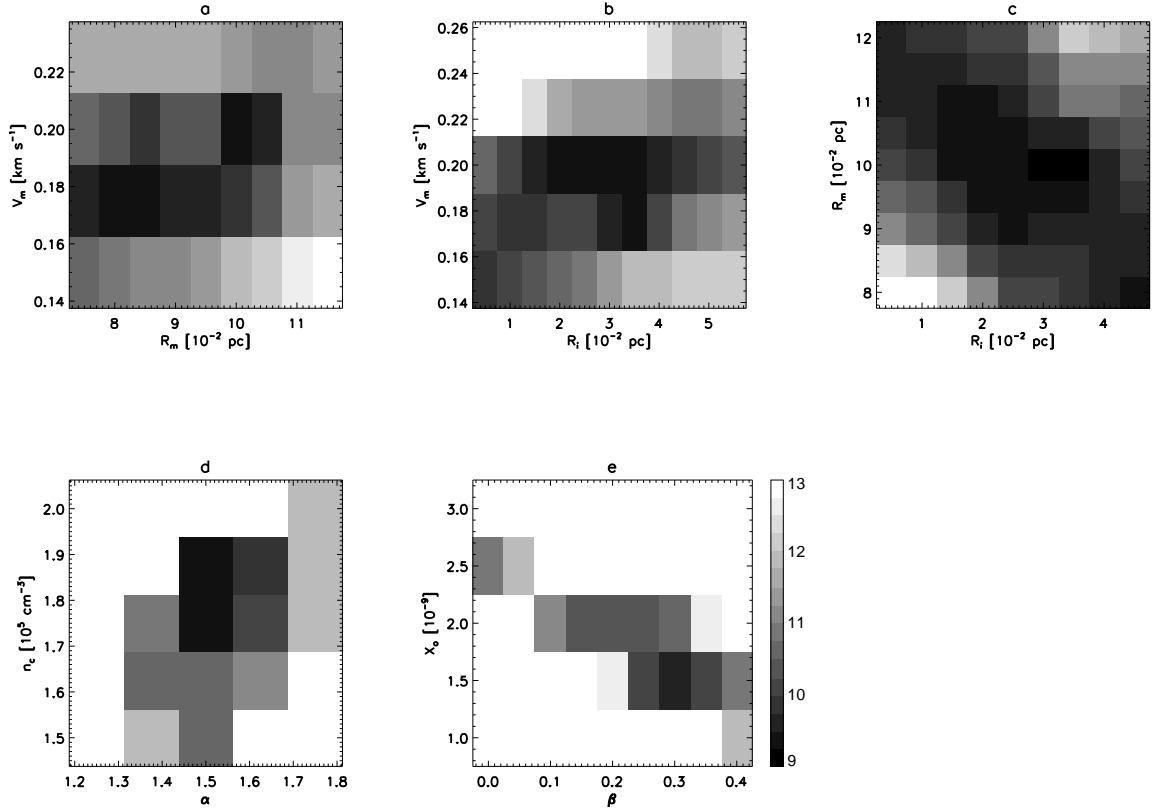}
\begin{center}
\caption{Distribution of $\chi^2$ around the best fit solution
with $\chi^2$ of 9.2 as a function of model parameters for L1197.
Others are the same as in Fig.~\ref{fig3}. \label{fig4}}
\end{center}
\end{figure}

The spectra of the best fit models are displayed in
Fig.~\ref{fig1}, where one can note that almost all the details
are reproduced, particularly for L694-2. The density and velocity
distribution of the  models are shown in Fig.~\ref{fig2}. The
minimum $\chi^2$'s are estimated to be 2.7 and 9.2 for L694-2 and
L1197, respectively. The distributions of the $\chi^2$ around the
best fit model are displayed in Fig.~\ref{fig3} and
Fig.~\ref{fig4}. As expected, the abundance parameters are not
completely independent, since both are related to the column
density of HCN. There seems to be (anti-)correlations among the
velocity parameters, too.


  To summarize, the infall velocity begins to increase from 0 near
the $R_i= 0.035$ pc for both the cores, and reaches the maximum
infall velocity of $V_{m}=  0.28$~km~s$^{-1}$ at $R_m = 0.085$~pc
for L694-2 and $V_{m}=  0.20$~km~s$^{-1}$ at $R_m = 0.10$~pc for
L1197 as shown in Fig.~\ref{fig2}.  The central density and power
index of L694-2 are around $n_c=1.75 \times 10^5$~cm$^{-3}$ and
$\alpha=2.5$, while those of L1197 are $n_c=1.75 \times 10^5 $~
cm$^{-3}$ and $\alpha=1.5$, respectively.
 Our result of L694-2 is similar to that derived from the near-infrared extinction and
 dust emission maps  \citep{harveya,harveyb}. The abundance is uniform with  $7\times 10^{-9}$ for
L694-2 ($\beta \approx 0.0$), while it increases slightly toward
the center for L1197 ($\beta \approx 0.3$) from $1.7 \times
10^{-9}$ to $3.5 \times 10^{-9}$.

\section{Discussion}
  Since we optimized several parameters cyclically, it is uncertain whether they are
   a set of best fit parameters.
  This may be true because we did not explore sufficient parameter spaces.
  However, since our optimization is based on the qualitative understanding
of line profile formation, it is unlikely that a quite different
solution will exists.
  The degree of coincidence of the synthesized and the observed line profiles also
supports this argument, as shown in
Fig.~\ref{fig1}.
  Although a new solution may be derived, but characteristic features such
as the $\Lambda$-shaped velocity field will remain unchanged.

 The fact that the velocity distribution is concentrated in the middle
 and outer layers is attributed to the selection of HCN hyperfine lines that can
 probe a wide range of core interiors and to the relaxation of the condition of a
 constant or monotonic decrease/increase in the velocity field.
  Our model naturally explains the ``extended inward motion'' observed for several
starless cores since the infall motion is dominant in the middle
and outer layers.
  The size of the infalling layer is $\approx 2 R_m \sim 0.2$ pc, which is
consistent with the observations \citep{myers}.
  The extended inward motion probed by CS toward L1544 and L1551 appears to be
exceptional in that it extends over very wide area.
  This may be due to other types of motion \citep{swift}.

  The velocity field that we found does not contradict the observation of
\citet{williams}, in which increasing inward motion toward the
center in the central 30$''$ region of L694-2 is derived. For
this, they used the optically thin N$_2$H$^+$ line to probe the
infall velocity of $<~ 0.1$~ km~s$^{-1}$ and therefore
  the more dominant motion in the middle and outer layers has been missed.
  In our case, we assumed that the velocity in the central region is zero in order to reduce the
  number of parameters. In fact, we have tested several sets of models whose velocity fields
  have non-zero values in the deeper part, and have found that the velocity of $<0.1$~ km~s$^{-1}$ is acceptable.

The distribution of the infall motion of our model is similar to
that of the ambipolar diffusion model and isothermal core collapse
model \citep{ciolek, foster} in the sense that the models also
have the $\Lambda$-shaped velocity field and their infall velocity
peaks lie just outside the central density-flat region. However,
the velocity gradient of our model inside the velocity peak
position is steeper than that predicted by the dynamic models. We
can observe such a sudden decrease in the infall motion in the
deeper layer in the so-called first collapse phase of the model
used in \citet{masunaga}. However, this occurs when the central
part is very dense and opaque at an evolutionary stage later than
that of our sample. Some mechanisms resisting against gravity such
as rotation may work; however, they are easily excluded since
rotation is very slow \citep{williams}. An oscillation of starless
cores around equilibrium may be able to explain this, as shown in
\citet{ketofield}, where pressure wave reflected from center
combined with inward motion due to perturbation from outside
results in the steep velocity gradient in the middle. However the
magnitude of the motion is much less than sound speed.  In the
case of L694-2 and L1197, the inward motion is similar to or
greater than the sound speed, suggesting that they are undergoing
dynamical collapse. The mechanism of a steeper pressure gradient
should be sought, but it is inappropriate to describe it in
greater detail at this stage.

The motion of the ambipolar diffusion model \citep{ciolek} is very
slow and therefore it cannot be used to explain the velocity field
derived in this study. This is because the model of \citet{ciolek}
is developed for L1544 when the infall velocity of this core was
known to be less than 0.1 km s$^{-1}$. If the magnetic field is
reduced to explain the infall motion of the cores with ambipolar
diffusion, the model will be supercritical and reduced to the
usual isothermal collapse model.

\section{Conclusion}
By combining high spatial resolution HCN observation toward L694-2
and L1197 with the Monte Carlo radiative transfer calculation, we
have found that the infall velocity is most dominant in the middle
and outer layers of the cores and the infall velocity can be as
large as 0.3~km~s$^{-1}$. The most important feature of our study
is that the velocity field is determined with such accuracy and
fidelity that it is possible to compare dynamics and observations
for the first time. An extensive investigation of the velocity
distribution in collapsing starless cores is in progress (Sohn et
al., in preparation). The result will clarify the dynamic status
of starless cores in greater detail.

\acknowledgements

This study was supported by the Ministry of Science and
Technology, Korea, under the grant R14-2005-058-01002-0. S.H.L.
used the PC cluster facility of Korea Astronomy and Space Science
Institute. Institute. C.W.L acknowledges support by KOSEF
R01-2003-000-10513-0 program.

\clearpage

\end{document}